\documentstyle[12pt]{article}

\begin{document}

{\bf On the effects of the vanishing of the non-metricity to the effective
string field equations}
\\

Abstract\\

Effective string field equations with zero-constrained torsion have been
studied extensively in the literature. But, one may think that the effects
of vanishing of the non-metricity have not been explained in detail and also
in according to some recent literature [4],[5], the action density in my
previous paper [3] is not complete. For these reasons, in this erratum, in
according to the effects of vanishing of the non-metricity to the field
equations, the action density of my previous paper [3] will be completed in
a variational setting. Furthermore, up to now, an unambiguous derivation of
the dilaton potential has not been given. If one thinks that vanishing of the
non-metricity gives a spontaneous breakdown of local Weyl invariance then a
dilaton potential is obtained unambiguously in this framework.

{\bf 1.Introduction and Motivation}

Effective string field equations with zero-constrained torsion have been
studied extensively in the literature. But, one may think that the effects
of vanishing of the non-metricity have not been explained in detail and also
in according to some recent literature [4],[5], the action density in my
previous paper [3] is not complete. For these reasons, in this erratum, in
according to the effects of vanishing of the non-metricity to the field
equations, the action density of my previous paper [3] will be completed in
a variational setting. Furthermore, up to now, an unambiguous derivation of
the dilaton potential has not been given. If one thinks that vanishing of the
non-metricity gives a spontaneous breakdown of local Weyl invariance then a
dilaton potential is obtained unambiguously in this framework.

{\bf 2.Theory}

In my previous paper [3], the first-order sector of the Lagrangian was
\begin{eqnarray}
\nonumber
{\cal{L}}=e^{-\phi}(R_{ab}\wedge^{\ast}e^{ab}+Tr(F\wedge^{\ast}F)
-\alpha d\phi \wedge^{\ast}d\phi+\beta H\wedge^{\ast}H)\\
\nonumber
+\gamma e^{-\frac{N-4}{N-2}\phi}R_{ab}\wedge R_{cd}\wedge^{\ast}e^{abcd} \\
+(dH-\epsilon R_{ab}\wedge R^{ab}-\epsilon^{\prime}Tr(F\wedge F))\wedge
\mu+T^{a}\wedge \lambda_{a}.
\end{eqnarray}
To handle zero-constrained torsion effectively, some Lagrange
multiplier $(N-2)-$ forms $\lambda_{a}$ are introduced. But, unfortunately
the vanishing of the torsion does not necessarily give the vanishing of the
non-metricity, namely, from Eq.(2.9) in [4]  
\begin{eqnarray}
D^{\ast}e^{a}=\;^{\ast}(e^{a}\wedge e_{b})\wedge T^{b}+4\;Q\wedge\;^{\ast}e^{a},
\end{eqnarray}
is satisfied, where $Q$ is a non-metricity object. For the vanishing of the
non-metricity, $f_{a}\;D^{\ast}e^{a}$ term will newly be added in the Eq.(1)
action density, where $f_{a}$ is a zero-form. The variation with respect to
$f_{a}$ gives $D^{\ast}e^{a}=0$, so $Q$, namely, non-metricity equals to zero.
The variation with respect to the $\omega$ has not been affected by this
$f_{a}\;D^{\ast}e^{a}$ term because the $\omega$ variation term,
$\omega^{ab}\wedge f_{a}\;^{\ast}e_{b}$ equals identically to zero.
So only the variation with respect to the $e^{a}$ term rests. This term is 
\begin{eqnarray}
\delta e^{a}\wedge\;^{\ast}(e^{c}\wedge e_{a})\wedge Df_{c}.
\end{eqnarray}
There is no any other restriction on $f_{c}$. So, as a simple solution,
if one selects $f_{c}=i_{c}dV(\phi )$ then the dilaton potential
$V(\phi )$ is obtained.

In my previous paper [3], in the cosmological solutions, $f_{c}$ has been
taken as zero.

[1].  T.Dereli, R.W.Tucker,Class.Q.Grav.{\bf 4} (1987) 791
 
[2].  T.Dereli, \c{S}. \"{O}zkurt, Class.Q.Grav.{\bf 8} (1991) 889 
 
[3].  \c{S}. \"{O}zkurt, Class.Q.Grav.{\bf 13} (1996) 265
  
[4].  T.Dereli, R.W.Tucker, JHEP 03 (2002) 041 

[5].  D.Grumiller, W.Kummer, D.V.Vassilevich, hep-th 04 (2002) 253

\end{document}